\documentstyle [12pt]{article}
\def\be{\begin{equation}}
\def\ee{\end{equation}}
\def\bea{\begin{eqnarray}}
\def\eea{\end{eqnarray}}
\def\pd{\partial}

\begin{document}

\begin{titlepage}

\title{A Discourse on the Benney Equation}

\author{ A.N. Leznov$^{1,2}$ and J.Sanchez Mondragon$^{1}$\\
\quad\\
$^{1}${\it Research Center on Engeneering and Applied
Science}\\
{\it Av. Universidad 1001 col. Chamilpa C.P. 62210}\\
{\it Cuernavaca Morelos Mexico}\\
$^{2}${\it  Institute for High Energy Physics, 142284
Protvino,}\\
{\it Moscow Region, Russia}}

\maketitle

\begin{abstract}

It is shown that the Benney system is equivalent to a one
body classical dynamical problem in which the interaction
potential
is a functional of the action function. The general solution
of this
equation is responsible for the general solution of the
Benney system.
Some particular solutions of  the selfconsistent
Hamilton-Jacobi
equation arising in this investigation are presented
in explicit form.

\end{abstract}
\end{titlepage}

\section{Introduction}

We use the form of the Benney \cite{BN} system from
\cite{TZ}, where it is
presented (`in its original form') as:
\be
\frac{\pd v}{\pd t}+v\frac{\pd v}{\pd x}-\left(\int^y_0 dy
\frac{\pd v}{\pd x}\right)
\frac{\pd v}{\pd y}+\frac{\pd h}{\pd x}=0,\quad \frac{\pd
h}{\pd t}+
\frac{\pd}{\pd x} \left(\int^h_0 dy v\right)=0 \label{benie}
\ee
where the unknowns are the functions $v\equiv v(t;x,y),\
h\equiv h(t,x)$.

This paper is not aimed at the problems of symmetry and
numerous reccurence relations which follow from the Benney
system  (\ref{benie}) and are described in \cite{BN, TZ, KM,
ZA, FAR, FF1, FF2} but directly to the problem of
construction of the general solution to it. The history and
literature on Benney system reader can be found in
\cite{TZ}.

The general solution in our approach to the problem can
provisionally be
divided into two steps. In the first  we propose a
parametric representation of the
solution of the Benney system (in implicit form) in terms of
an assumed
distribution function. In the second step we propose a self-
consistent equation, which this function should satisfy. The
general solution
of this equation is responsible for the general solution of
the Benney system.

Thus the goal of this paper is to demonstrate that the
Benney system can be
presented in the form of one body classical dynamic
Hamilton-Jacobi equation,
with the potential expressed nonlocally in terms of the
action function.
We discuss also the connection of this problem with an
ordinary
differential equation of the second order. A particular
solution of (\ref{benie}),
depending on two arbitrary functions each of one argument is
constructed.
However the general solution of this equation must depend on
one arbitrary
function of two independent arguments and one function of
one argument
(the initial values of the functions $v,\ h$). Such a
solution is unknown
for us at this moment.

The paper is organized in the following way. In section 2 we
consider in detail
the case  with $h=0$ and explain the strategy of the
calculations, which  are also
applicable to the general case without any considerable
change. In sections
3 and 4 the general case is considered and shown that the
Benney system is
equivalent to a single non-linear integro-differentional
equation, which in
turn is equivalent to a Hamilton-Jacobi one, with the
potential in the form of
a nonlocal functional of the second order derivatives of the
action function. In
terms of the general solution (not only of its first
integral!) of the self-
consistent Hamilton-Jacobi equation the general solution of
the Benney system
(\ref{benie}) is represented in implicit form. Two
different particular solutions of this equation with
functional arbitrariness
of two arbitrary single argument functions are also
presented here.
The connection of the self-consistent Hamilton-Jacobi
equation with an ordinary
differential equation of the second order and the ways of
obtaining its
general solution are discussed in section 5. In section 6 we
represent our
results in the form of a theorem. Section 7 is devoted to a
discussion of the
results of the paper and discussion of possible perspectives
for future
investigation. In the Appendix the reader will find the
connection of the
solution of the main equation of the second subsection of
section 3 with the
Hamilton-Jacobi equation in a field of force depending only
upon the time
variable.

\section{Preliminary manipulations}

This section contains   detailed calculations leading to  a
parametrical
representation of the general solution of the Benney system
in implicit form
under the simplest assumption that $h=0$ in (\ref{benie}).
The reader for whom the result is more interesting than its
derivation can begin
reading directly the parametrical representation of the
solution
(\ref{BPP}), which is checked independently in the next few
lines.

Let us define $u(t,x,y)\equiv\int^y_0 dy' v(t,x,y')$  and
resolve the second
equation of the system (\ref{benie}) as
\be
h=H_x(t,x),\quad u(t;x,H_x)=-H_t \label{HD}
\ee

After such a substitution, the system (\ref{benie}) takes
the form:
\be
u_{t,y}+u_yu_{xy}-u_xu_{yy}+H_{xx}=0,\quad
u(t;x,H_x)=-H_t,\quad u(t;x,0)=0
\label{b1}
\ee

To have an experience in working with the last system of
equations let us at
first restrict ourselves to the particular solution for
which $H=Const$ (really
this is an inessential restriction for the parametrical
representation of the
solution as can be seen from the results of the section 3).

Thus we have to solve the single equation for the unknown
function $u$ with the
boundary condition:
\be
u_{t,y}+u_yu_{xy}-u_xu_{yy}=0,\quad u(t;x,0)=0 \label{b1s}
\ee

Let us present the solution of (\ref{b1s}) in the form
$\phi(u;t,x,y)=$const,
considering $\phi$ as an unknown function with four
arguments.
Direct calculations of the first and second order
derivatives using the rules of
differentiation of implicit  functions leads to the result:
\be
Det_3 \pmatrix{0 & \phi_u & \phi_t \cr
             \phi_u & \phi_{uu} & \phi_{ut} \cr
             \phi_y & \phi_{uy} & \phi_{yt} \cr} +
             Det_3 \pmatrix{0 & \phi_y & \phi_x \cr
             \phi_y & \phi_{yy} & \phi_{yx} \cr
             \phi_u & \phi_{uy} & \phi_{ux} \cr}=0
             \label{b2}
\ee

The following notation will be used:
\be
\alpha={\phi_t\over \phi_u},\quad \beta={\phi_x\over
\phi_u},\quad\lambda=
{\phi_y\over \phi_u}\label{NN}
\ee
$$
\hat A=\frac{\partial}{\partial y}-\lambda
\frac{\partial}{\partial u},\quad
\hat B=\frac{\partial}{\partial x}-\beta
\frac{\partial}{\partial u},
\quad \hat C=\frac{\partial}{\partial t}-\alpha
\frac{\partial}{\partial u}
$$
The system of equations,  which the functions $\alpha,\beta,
\lambda$ satisfy (as a consequence of their definition only)
takes the form:
\be
[\hat A,\hat B]=[\hat B,\hat C]=[\hat C,\hat A]=0\label{MS}
\ee
In addition directly from (\ref{b2}) the main equation for
them,
which can be written in two equivalent forms, follows:
\be
\{\phi,\lambda\}_{tu}=\{\phi,\lambda\}_{xy},\quad
\{\phi,\alpha\}_{yu}=\lambda^2
\{\phi,{\beta\over \lambda}\}_{yu}\label{BUIT}
\ee
where $\{X,Y\}_{a,b}\equiv X_aY_b-X_bY_a$. Considering
$\alpha,\beta$ as
functions depending upon four independent arguments
$\phi,\lambda,x,t$, we
obtain for the last equation of (\ref{BUIT}):
\be
(\lambda_y - \lambda \lambda_u) (\alpha_{\lambda} -\lambda^2
({\beta\over \lambda})_{\lambda}) = 0\label{BUITI}
\ee
and rewrite (\ref{MS}) as:
\be
(\lambda_t-\alpha
\lambda_u)\beta_{\lambda}+\beta_t=\alpha_x+(\lambda_x-\beta
\lambda_u)\alpha_{\lambda}\label{MFS}
\ee
$$
(\lambda_x-\beta \lambda_u)=(\lambda_y-\lambda
\lambda_u)\beta_{\lambda},\quad
(\lambda_y-\lambda
\lambda_u)\alpha_{\lambda}=(\lambda_t-\alpha \lambda_u)
$$
Combining the first equation of (\ref{MFS}) with the last
two we
conclude that:
\be
\alpha_x=\beta_t,\quad \alpha=\Theta_t,\quad \beta=\Theta_x
\label{GR}
\ee
and after substitution of(\ref{CR}) into (\ref{BUITI}) we
arrive at a linear
equation for a single function $\Theta$:
\be
-\Theta_x+\lambda
\Theta_{x,\lambda}=\Theta_{t,\lambda}\label{BUITII}
\ee

The most straightforward way to obtain its solution is by
differentiation  with
respect to the argument $\lambda$. This leads it to:
$$
(\frac{\partial}{\partial t}-\lambda
\frac{\partial}{\partial x})\Theta_{
\lambda,\lambda}=0
$$
The general solution of the last equation is obvious:
$$
\Theta=\int^{\lambda} d \lambda' (\lambda-\lambda')
G(x+\lambda't,\lambda';\phi)
+\lambda K(x,t;\phi)+L(x,t;\phi)
$$
The initial equation (\ref{BUITII}) connects the functions
$K$ and $L$ by a
gradient dependence and, finally, solution of
(\ref{BUITII}), depending
on two arbitrary functions $G,F$ (each with three
independent arguments) takes
the form:
\be
\Theta=\int^{\lambda} d\lambda' (\lambda-\lambda')
G(x+\lambda't,\lambda';\phi)+
F_t-\lambda F_x \label{NF}
\ee

In what follows we will denote the "moments" of the function
$G$ by the
corresponding upper index:
$$
G^s(x,t;\phi,\lambda)=\int^{\lambda} d\lambda' (\lambda')^s
G(x+\lambda't,
\lambda';\phi)
$$
In this notation $\alpha$ and $\beta$ functions take the
form:
$$
\alpha=(F_{tt}-G^1_t)-\lambda (F_{xt}-G^0_t),\quad
\beta=(F_{xt}-G^1_x)-\lambda (F_{xx}-G^0_x)\label{FNF}
$$

Substituting $\alpha,\beta,\lambda$ via derivatives of the
function $\phi$  from
(\ref{NN}) we come to the following system of equations of
the first order:
\be
\phi_x=(F_{xt}-G^1_x)\phi_u-(F_{xx}-G^0_x)\phi_y,\quad
\phi_t=(F_{tt}-G^1_t)
\phi_u-(F_{xt}-G^0_t)\phi_y \label{NNF}
\ee
Up to now we have used  only one equation from the system
(\ref{MFS}) which
together with the second equation (\ref{BUIT}) leads to
(\ref{FNF}).
Two remaining equations (\ref{MFS}) have as their direct
corollary the following
system:
\be
\lambda_x = (F_{xt} - G^1_x)\lambda_u - (F_{xx} -
G^0_x)\lambda_y, \quad \lambda_t = (F_{tt}-G^1_t) \lambda_u
- (F_{tx} - G^0_t)\lambda_y \label{NNF1}
\ee
The systems (\ref{NNF}) and (\ref{NNF1}) are selfconsistent
in the sense of
equality of the second mixed derivatives and our nearest
goal now is to
extract the consequences, which follow from them.

Let us consider in (\ref{NNF}) the function $\phi$ depending
on two sets of
four arguments $\phi\equiv
\theta(x,t,\lambda;u)=\psi(x,t,\lambda;y)$. Keeping in
mind (\ref{NNF1}) we can transform (\ref{NNF}) to two forms:
$$
\theta_x=(F_{xt}-G^1_x)\theta_u,\quad
\theta_t=(F_{tt}-G^1_t)\theta_u,
$$
$$
\psi_x=-(F_{xx}-G^0_x)\psi_y,\quad
\psi_t=-(F_{xt}-G^0_t)\psi_y
$$
with the obvious general solution:
\be
\phi = \theta = P(u + F_t - G^1, \lambda) = Q(y - F_x +
G^0,\lambda) \label{PB}
\ee
where $P$ and $Q$ are  arbitrary functions of their two
arguments.
Resolving (\ref{PB}) with respect to the first arguments of
$P$ and $Q$
functions, we obtain a system of two functional equations,
which the
functions $\phi$ and $\lambda$ satisfy:
\be
u+F_t-G^1=R(\lambda,\phi),\quad
y-F_x+G^0=S(\lambda,\phi)\label{PPB}
\ee
Here $R$ and $S$ are  arbitrary functions of two arguments
(inverses of
 $P,Q$).

We recall that $F=F(x,t;\phi),\  G^0(x,t,\lambda,\phi),\
G^1(x,t,\lambda,\phi)$
and for this reason (\ref{PPB}) is a system of functional
equations
which determine the functions $\lambda,\phi$ implicitly. By
the rules of
differentiation of  implicit functions it may be seen that
the functions $\lambda$
and $\phi$ defined by (\ref{PB}),(\ref{PPB}) satisfy
the systems (\ref{NNF}) and (\ref{NNF1}).

To show this and to find the additional condition which
connects the
functions $R,S$ (arbitrary up to now) let us calculate
derivatives of
the functions $\lambda,\phi$. For this the following
abriviations will be used
$$
A=R_{\phi}+G^1_{\phi}-F_{t,\phi},\quad
B=R_{\lambda}+G^1_{\lambda},\quad
C=S_{\phi}-G^0_{\phi}+F_{x,\phi},\quad
D=S_{\lambda}-G^0_{\lambda}
$$
and the linear system of equations from which the
corresponding derivatives are
determined takes the form:
$$
F_{tt}-G^1_t=A\phi_t+B\lambda_t, \quad
-F_{xt}+G^0_t=C\phi_t+D\lambda_t,
$$
$$
F_{tx}-G^1_x=A\phi_x+B\lambda_x,\quad
-F_{xx}+G^0_x=C\phi_x+D\lambda_x,
$$
$$
1=A\phi_u+B\lambda_u,\quad 0=C\phi_u+D\lambda_u,
$$
$$
0=A\phi_y+B\lambda_y,\quad 1=C\phi_y+D\lambda_y.
$$
One can verify that the linear systems (\ref{NNF}) and
(\ref{NNF1}) are
satisfied and for $\lambda$ we obtain:
$$
\lambda={\phi_y\over \phi_u}=-{B\over
D}=-{R_{\lambda}+G^1_{\lambda}\over
S_{\lambda}-G^0_{\lambda}}
$$
{}From the last expression, keeping in mind the definitions
of $G^0$ and $G^1$
functions it immediately follows that:
$$
\lambda S_{\lambda}=-R_{\lambda}
$$
The last equation may be resolved in terms of only one
arbitrary function $T$ of
two arguments $\lambda$ and $\phi$:
$$
R=\lambda T_{\lambda}-T,\quad S=-T_{\lambda}
$$

To come back to the solution of the initial system
(\ref{b1}) and express $u$ as
a function of the arguments $(x,t,y)$ it is necessary to put
$\phi=$constant in
all formulae above. After such a substitution the functions
$G$ and $F$ reduce
 to  functions of two indepentent arguments and $T$  to only
 one,
$\lambda$.

Substituting the last expressions into (\ref{PPB}) and
recalling the
boundary condition ($H=$constant) from we obtain:
\be
u+F_t-G^1=\lambda T_{\lambda}-T,\quad
y-F_x+G^0=-T_{\lambda},\quad,
u(x,t,0)=0\label{PPBB}
\ee

It is not difficult to check that the function $T$ can be
included in
$G$ by the of substitution:
$$
G(x_1,x_2)\to G(x_1,x_2)+T_{x_2,x_2}(x_2),
$$
which is equivalent to equating $T$ in (\ref{PPBB}) to zero.

Thus finally (\ref{PPBB}) takes the form:
\be
u+F_t-G^1=0,\quad y-F_x+G^0=0,\quad u(x,t,0)=0\label{PBP}
\ee

Two first equations define parametrically (via
$\lambda=-u_y=-v$) $u$ as
function of its three arguments. The third one can be
considered as a boundary
condition, but really this is an equation  determining two
two dimensional
functions $F(x,t),\lambda(x,t,0)\equiv \nu$. Indeed,
substituting $u=y=0$ into
(\ref{PBP}) we obtain:
\be
F_t-G^1(x,t,\nu)=0,\quad -F_x+G^0(x,t,\nu)=0\label{OK}
\ee
Comparing the second mixed partial derivatives of the
function $F$
and keeping in mind the definition of the "moments"
$G^1,G^0$
we obtain finally:
\be
(\nu_t-\nu\nu_x)G(x+\nu t,\nu)=0\label{OOKK}
\ee
It is easy to check, that the solution of Monge equation:
$$
\nu_t-\nu\nu_x=0
$$
is implicitly defined by the equation
$$
G(x+\nu t,\nu)=g={\rm constant}
$$

Below we rewrite (\ref{PPBB}) including the function $F$  in
the lower limit of
the integrals and emphasise the functional dependence of all
functions involved:
\be
u(t,x,y)-\int^{-v(t,x,y)}_{\nu(t,x)} d\lambda \lambda
G(x+\lambda t,\lambda)=0
\quad
y+\int^{-v(t,x,y)}_{\nu(t,x)} d\lambda G(x+\lambda
t,\lambda)=0 \label{BPP1}
\ee
$$
\nu_t-\nu\nu_x=0
$$

The equations (\ref{BPP1})  define implicitly the general
solution of
the Benney system in the case $h=0$, depending on one
function $G$ of two
independent arguments (which is in  connected with the
initial value of the
function  $v$).

\section{The general case of the Benney equation}

Now we consider the general case of the  Benney system
(\ref{benie}). All
steps of the calculations will be the same as the previous
with the additional
terms connected with $H_{xx}$ in (\ref{b1}).

Equation (\ref{b1s}) now reads:
\be
Det_3 \pmatrix{0 & \phi_u & \phi_t \cr
             \phi_u & \phi_{uu} & \phi_{ut} \cr
             \phi_y & \phi_{uy} & \phi_{yt} \cr} +
             Det_3 \pmatrix{0 & \phi_y & \phi_x \cr
             \phi_y & \phi_{yy} & \phi_{yx} \cr
             \phi_u & \phi_{uy} & \phi_{ux}
             \cr}+H_{xx}\phi^3_u=0 \label{b3}
\ee
The definitions of the functions $\alpha,\ \beta,\ \lambda$
(\ref{NN}) do not change
and the equations which follow from (\ref{MS}), (\ref{MFS})
are preserved in form.

As a corollary of (\ref{b3}) the modified main equation
(\ref{BUITI}) is:
\be
(\lambda_y - \lambda\lambda_u) (\alpha_{\lambda} - \lambda^2
({\beta \over \lambda})_{\lambda}) + H_{xx} = 0
\label{BUITs}
\ee

The equation for the function $\Theta$ takes the form:
\be
(\lambda_y-\lambda\lambda_u)(\Theta_x-\lambda
\Theta_{x,\lambda}+
\Theta_{t,\lambda})+H_{xx}=0\label{BUITIs}
\ee

Further calculations word for word repeat the same story of
the previous section.
The crucial point is the understanding that an arbitrary
function of four
arguments $\Theta\equiv \Theta(t,x,\lambda,\phi)$ may be
presented in the form:
\be
\Theta=\int^{\lambda} d\lambda' (\lambda-\lambda')
G(t,x,\lambda';\phi)+
F(t,x,\phi)+\lambda \Phi(t,x,\phi) \label{NNFF}
\ee
where $G=\Theta_{\lambda,\lambda}$.
Indeed (\ref{NNFF}) is one  possible form a Taylor series
for the function $\Theta$
with respect to one of its independent arguments. The form
(\ref{NNFF})
doesn't  have any connection with equation which the
function $G$
satisfy.

All formulae between (\ref{NNF}) and (\ref{OK}) preserve
their form with the
obvious exchange  $-F_x\to \Phi,F_t\to F$ and lead to
the final result; instead of (\ref{OK}):
\be
u+F-G^1=0,\quad y+\Phi+G^0=0,\quad u(x,t,0)=0,\quad
u(x,t,H_x)=-H_t,
\label{BP}
\ee
where $F,\Phi$ are two arbitrary functions of three
arguments $(t,x,\phi)$.

After differentiation of the second equation (\ref{BP}) with
respect to
operator $\frac{\partial}{\partial y}-\lambda
\frac{\partial}{\partial u}$
we obtain:
$$
(\lambda_y-\lambda\lambda_u)=-{1\over
\Theta_{\lambda,\lambda}}
$$
Substituting the last expression into (\ref{BUITIs}) and
subsequent
differentiation of the result with the respect to the
argument $\lambda$  we obtain
an equation for function $G\equiv \Theta_{\lambda,\lambda}$:
\be
-G_t+\lambda G_x-H_{xx} G_{\lambda}=0\label{G}
\ee
The last equation is exactly the equation for the
distribution function for  a one
body classical dynamical system with Hamiltonian function
${1\over 2}\lambda^2+
H_x$. Indeed, the condition of conservation of some quantity
along the trajectory can
be written as:
$$
\rho_t=\{H,\rho\}_{\lambda,x}
$$
This is exactly the equation for the function  $G$ above. (
In the quantum
case it is the equation for the density matrix with the
exchange of Poisson
brackets with commutators).

Now let us include functions $F,\Phi$ from (\ref{BP}) into
the first boundary
condition and rewrite (\ref{BP}) in the form:
\be
u-\int^{-v}_{\nu} d\lambda \lambda G(t,x,\lambda)=0\quad
y+\int^{-v}_{\nu} d\lambda G(t,x,\lambda)=0 \label{BPP}
\ee
where $\nu$ is a function of two arguments $t,x$.It is
obvious that to satisfy
the first boundary condition $\nu=-v(t,x,0)$. \footnote{We
remind the reader
that in all formulae below it is necessary to put
$\phi=$constant, as
explained in the previous section.}

The condition on the second boundary leads to the additional
restiction:
\be
H_t+\int^{\mu}_{\nu} d\lambda \lambda G(t,x,\lambda)=0\quad
H_x+\int^{\mu}_{\nu} d\lambda G(t,x,\lambda)=0 \label{BBPP}
\ee
($\mu=-v(x,t,-H_x)$).

Now let us check by direct calculation the conditions under
which functions
$u(t,x,y),\ H(t,x)$ defined by the (\ref{BPP}) and
(\ref{BBPP}) are a solution of
the Benney equation (\ref{b1}).

In the notation used above the Benney equation may be
rewritten
as ($u_y=v=-\lambda$):
$$
v_t+vv_x-u_xv_y+H_{xx}=0
$$
After differentiation of the second equation (\ref{BPP})
with respect to $t$
we find:
$$
-v_tG(t,x,-v)-\nu_tG(t,x,\nu)+\int^{-v}_{\nu} d\lambda
G_t(t,x,\lambda)=0
$$
Substituting the derivatives of $G_t$ from the equation for
it (\ref{G})
we obtain finally:
$$
-v_t=-{\int^{-v}_{\nu} d\lambda \lambda
G_x(t,x,\lambda)-H_{xx}
(G(-v)-G(\nu))-\nu_tG(\nu)\over G(-v)}
$$

By the same technique we have:
$$
u_xv_y={\int^{-v}_{\nu} d\lambda \lambda G_x(t,x,\lambda)+
(vv_xG(-v)-\nu\nu_xG(\nu))\over G(-v)}
$$
or
$$
-(vv_x-u_xv_y)={\int^{-v}_{\nu} d\lambda \lambda
G_x(t,x,\lambda)-\nu\nu_xG(\nu)\over G(-v)}
$$
After summation of these results we obtain the equality:
$$
v_t+vv_x-u_xv_y=-H_{xx}+{G(\nu)\over G(-v)}
(\nu_t-\nu\nu_x-H_{xx})
$$
and conclude that in order to satisfy the Benney equation
$\nu$ function must be the
solution of the equation:
$$
\nu_t-\nu\nu_x-H_{xx}=0
$$

The condition of equality of the second mixed partial
derivatives
with respect to $t,x$ of the function
$H$ follows from (\ref{BBPP}) and leads to the
conclusion that the function  $\mu$ satisfies the same
equation as $\nu$
and the condition of selfconsistency of the whole
construction my be written
in the form:
\be
H_{xx}=\nu_t-\nu\nu_x=\mu_t-\mu\mu_x,\quad
H_x=\int^{\mu}_{\nu} d\lambda
G(t,x,\lambda)\label{FV}
\ee
where $G$ satisfies (\ref{G}).

Considering $\lambda$ in equation (\ref{G}) as an unknown
function of three
independent arguments $(x,t,g)$ and defining:
$$
g=G(x,t,\lambda)
$$
we arrive at  the equation, which this function satisfies:
\be
\lambda_t-\lambda\lambda_x-H_{xx}=0\label{XY}
\ee
from which it is follows that:
$$
\nu=\lambda(x,t,g_1),\quad \mu=\lambda(x,t,g_2)
$$
After exchaging  integration variables under the integral
sign in
(\ref{BBPP}) we amalgamate the Benney equation with the
corresponding boundary
conditions into a single (seemingly very strange)
integro-differential
equation:
\be
\lambda_t-\lambda\lambda_x-\int^{g_2}_{g_1} g dg
\lambda_{g,x}=0\label{CR}
\ee
The main function $G$ from (\ref{G}) may be reconstructed
implicitly
via the solution of the last equation:
$$
\lambda=\lambda(x,t,G)
$$

\subsection{The simplest particular solution of the Benney
system}

In this subsection we would like to demonstrate the
selfconsistency of
the previous construction by the example of the "simplest"
solution
of Benney system. This also will give us the possibility of
understanding which
class of functions is typical for this problem.

Of course, distribution function $G=-{1\over 2A}=$ constant
satisfies (\ref{G})
with an arbitrary choice of the potential function $H_x$.
Thus as a corollary of
(\ref{BBPP}) we have ($G=-{1\over 2A}$):
$$
H_x={1\over 2A} (\mu-\nu),\quad H_t={1\over 4A}
(\mu^2-\nu^2)
$$

{}From (\ref{BPP}) we obtain:
\be
v=-2Ay-\nu,\quad u=-Ay^2-\nu y, \quad h=H_x={1\over 2A}
(\mu-\nu)  \label{EH1}
\ee
By direct computation one can become convinced that
(\ref{EH1})
is really the solution of the Benney system if $\mu,\nu$
functions satisfy
(\ref{EH}), which in the case under consideration takes the
form
\be
\mu_t-\mu\mu_x=\nu_t-\nu\nu_x={1\over 2A} (\mu-\nu)_x
\label{EH}
\ee

For linearization of the last system let us exchange  the
dependent and independent variables. In other words let us
consider $(t,x)$
as a functions of the pair of variables $(\mu,\nu)$.
Corresponding formulae for
this transformation are obvious:
$$
t=t(\mu,\nu)\quad 1=t_{\nu}\nu_t+t_{\mu}\mu_t,\quad
0=t_{\nu}\nu_x+t_{\mu}\mu_x
$$
$$
x=x(\mu,\nu)\quad 0=x_{\nu}\nu_t+x_{\mu}\mu_t,\quad
1=x_{\nu}\nu_x+x_{\mu}\mu_x
$$
After solving of the last equations and substitution of the
result into
(\ref{MO}) we obtain the first order linear system of
equations:
$$
(x+{\nu}t)_{\mu}=-(x+{\mu}t)_{\nu}={1\over 2A}
(t_{\mu}+t_{\nu})
$$
{}From the last system we conclude that
$$
(x+{\nu}t)=-\theta_{\nu},\quad (x+{\mu}t)=\theta_{\mu},
$$
\be
t={\theta_{\nu}+\theta_{\mu}\over {\mu}-{\nu}},\quad
x=-{\mu\theta_{\nu}+
\nu\theta_{\mu}\over {\mu}-{\nu}}\equiv {1\over
2}(\theta_R-{\Sigma\over
R}\theta_{\Sigma}),\label{REZ}
\ee
and obtain the equation, which the function $\theta$
satisfy. We present it
in terms of the variables $\Sigma={1\over
2}(\mu+\nu),R={1\over 2}(\mu-\nu)$
in which it takes the most simple form
$$
(1-{1\over AR})\theta_{\Sigma,\Sigma}=\theta_{R,R}
$$

This is a linear equation of  second order with the
separable variables.
It allows us to obtain its general solution depending on two
arbitrary functions, each of one argument. Equations
(\ref{REZ}) have to be
reversed and variables $\mu,\nu$ may be expressed as
functions of the initial
variables $x,t$.

Formulae (\ref{EH1}) give a paricular solution of the Benney
system depending on
two arbitrary functions each of one argument. Of course,
this is
not the general solution of this system which must depend on
one function of
two arguments ( the initial value $v(x,y,0)$) and one
function of one
argument (the initial value  $h(x,0)$).

The example considered  shows that it is impossible to
expect some simple analytical expression for the general
solution of
the Benney system.

\subsection{Particular solution of main equation for
$\lambda$ function}

 Let us seek the solution of main equation (\ref{CR}) in the
 form:
$$
\lambda=x A(t,g)+B(t,g)
$$
We have in consequence:
\be
(x A_t+B_t)-A(x A+B)+(\int^{g_2}_{g_1} dg g (x
A_g+B_g))_x=0\label{FVF}
\ee
Equating coefficients at zero and unity, decreasing x to
zero and solving
the system of equations arising  for the functions $A,B$ we
obtain finally:
\be
\lambda=-{x+\Phi(t)+V(g)\over t+U(g)}+\Phi_t,\quad
\Phi=\int^{g_2}_{g_1} dg g
U_g\ln (t+U(g)) \label{L}
\ee
where $U(g),V(g)$ are arbitrary functions.

Resolving (implicitely) the equation (\ref{L}) with respect
to the function
$G$$$
\lambda=\lambda (x,t,G)
$$
and substituting the result into (\ref{BPP}) we come to the
solution of the Benney
system depending on two arbitrary functions $U,V$, each of
one argument. Of
course this solution is of a different kind to that in the
previous
subsection. In this case the distribution function $G$ is
different from a
constant. In some cases under a definite choice of the  form
$U,V$ functions
all calculations may be done explicitely and solution of
Benney system can be
presented by analytical expressions.

In the Appendix the reader can find  another way to obtain
the solution of the
present subsection, connected with the ordinary differential
equation of the
second order (see section 5).

\section{One body classical mechanical problem with
self-consistent
potential of interaction}

Equation (\ref{CR}) has the form of two dimensional
conservation law. It
allows us to solve this equation in the form:
$$
\lambda=-S_x,\quad {\lambda^2\over 2}+\int^{g_2}_{g_1} g dg
\lambda_g=-S_t
$$
After eliminating the function $\lambda$ we pass to the one
body Hamilton-Jacobi equation:
\be
S_t+{S_x^2\over 2}-\int^{g_2}_{g_1} g dg S_{g,x}\equiv
S_t+{S_x^2\over 2}+
V(x,t)=0,\label{HJ}
\ee
where the potential function $V$ is expressed nonlocally via
the derivatives
of action function $S$.

We would like to point out here that a situation of this
kind is not
a new one. It is sufficient to remember the famous
non-linear Schr\"odinger
equation, where the potential of interaction coincides with
the square
of the modulus of the wave function. In the problem
considered here
the action function  plays the same role  (classical limit
of the wave function).
The potential function is expressed in a
non-local way in terms of the derivatives of this
(\ref{HJ}).

\section{Connection with a second order ordinary
differential equation}

After differentiation of the main equation (\ref{CR}) with
respect to
the argument $g$
(keeping in mind that the integral term of equation doesn't
depend upon
$g$), we obtain:
$$
(\lambda_g)_t=(\lambda\lambda_g)_x,\quad \lambda_g=f_x,\quad
\lambda\lambda_g=
f_t
$$
where the function $f$ satisfies in turn the equation
arising after
eliminating the function $\lambda$  from the previous
system:
$$
({f_t\over f_x})_g=f_x
$$
Considering in the last equation $x\equiv X$ as an unknown
function of the
independent arguments $(t,f,g)$ we have in consequence:
\be
f=f(X,t,g),\quad 1=f_x X_f,\quad 0=f_t+f_x X_t,\quad
0=f_g+f_x X_g\label{DDX}
\ee
and as a corollary the equation, which the function $X$
satisfies:
\be
X_g X_{t,f}-X_f X_{t,g}=1\label{!}
\ee
Differentiation of the last equation with respect to the
argument $t$ leads to:
$$
X_{ttf}X_g- X_{ttg}X_f=0
$$
This means that the function $X_{tt}$ depends on the
arguments $(g,f)$ only via
the function $X$  and thus we have
\be
X_{tt}=Q(X,t)\label{!!a}
\ee
In other words, from (\ref{!}) it follows that function $X$
is the solution
of a ordinary differential equation of the second order
(\ref{!!a}) with
respect to the argument $t$.

Let us present the general solution of this equation in the
form:
$$
X=X(c_1,c_2;t)
$$
and fix the choice of integration $c_1,c_2$ constants by the
condition
\footnote{Let us recall that these constants are  defined up
to a
general covariant transformation. This circumstance always
allows us to satisfy the
condition below.}:
$$
X_{c_1} X_{t,c_2}-X_{c_2} X_{t,c_1}=1
$$
Then $c_1,c_2$ as functions of the arguments $(f,g)$ thanks
to (\ref{!})
are connected by a canonical transformation with the
generating function
$W(c_1,g)$:
$$
c_2=W_{c_1}(c_1,g),\quad f=W_g(c_1,g)
$$

Collecting all these results we are able to rewrite the main
equation
(\ref{CR}) in the form ofa functional integral equation
determining  the
unknown function $X$:
$$
x=X(W_{c_1},c_1,t),\quad X_{c_1} X_{t,c_2}-X_{c_2}
X_{t,c_1}=1,
$$
\be
X_{tt}(W_{c_1},c_1,t)-(\int^{g_2}_{g_1} dg g
W_g)_{xx}=0,\quad
c_2=W_{c_1}(c_1,g)\label{VBNF}
\ee
where, as  was mentioned above, $W(c_1,g)$ is the generating
function of the
canonical transformation.

In (\ref{VBNF}) the unknowns are both $X$ and $W$ and we
have no idea
at this moment of any constructive solution of this system,
except by
guesswork. In the  Appendix we show that the choice of the
second
order ordinary differential equation in the form
$X_{tt}=Q(t)$ corresponds
to a solution of  subsection 2 in the section 3.

\section{The main Theorem}

The results of the previous sections may be summarised in
the following

Theorem

Each solution of nonlinear integro-differential equation
$$
\lambda_t-\lambda\lambda_x-\int^{g_2}_{g_1} dg g
\lambda_{g,x}=0
$$
is connected (implicitly) with the solution of Benney system
$$
\frac{\pd v}{\pd t}+v\frac{\pd v}{\pd x}-\left(\int^y_0 dy
\frac{\pd v}{\pd x}\right)
\frac{\pd v}{\pd y}+\frac{\pd h}{\pd x}=0,\quad \frac{\pd
h}{\pd t}+
\frac{\pd}{\pd x}\left(\int^h_0 dy v\right)
=0
$$
via the following formulae
$$
u(x,y,t)-\int^{-v(y,x,t)}_{\nu(x,t)} d\lambda \lambda
G(t,x,\lambda)=0\quad
y+\int^{-v(y,x,t)}_{\nu(x,t)} d\lambda G(t,x,\lambda)=0
$$
$$
u(x,y,t)=\int^y_0 dy'v(y',x,t),\quad h=\int^{\mu}_{\nu}
d\lambda
G(t,x,\lambda)\equiv \int^{g_2}_{g_1}dg g\lambda_g
$$
where
$$
\mu=\lambda(t,x,g_2),\quad\nu=\lambda(t,x,g_1),\quad
\lambda=\lambda(t,x,G)
$$

\section{Outlook}

The main result of the paper consists in the Theorem of the
previous section,
giving asolution of the Benney system parametrically in
implicit form
in terms ofa  distribution function and a selfconsistent
equation (\ref{CR}),
which this function must satisfy.
If it is possible to find the general solution of the last
equation in the future
the problem of  the construction of the general solution of
Benney system would be
solved with the help of formulae (\ref{BBPP}).

The most interesting questions about possible
representations of the results
in the form available for experimental application (the
origin of the Benney
system is the hydrodynamical problem of  surfaces waves
\cite{BN}) are out of
the framework of the present paper. To encompass these
questions will be
possible after more detailed investigation of the main
equation (\ref{CR}) and
the discovery of ways of its regular integration (maybe on
the level of the
computer computations).

All results of the present paper are obtained on intutive
background and
demand for the their rigorous foundation more powerfull
mathematical methods.
The authors can guess that they connected with the better
known investigation
of the inner symmetry group of Benney system. We hope to
come back to these
interesting questions in our future publications.

\section{Acknowledgements.}

Autrhors thanks D.B.Fairlie, S.F.Luzanov and I.D.Pleshakov
for discussion
of the results and big help in the preparation the
manuscript.

\section{Appendix}

In this Appendix we would try to widen the solution of the
second subsection
and obtain it on a more systematic background. With this
purpose let us
consider more precisely and evaluate the integral in
(\ref{VBNF}).
\be
f_{xx}=({1\over X_f})_x=-{X_{ff}\over X_f^2}f_x={1\over
2}({1\over X_f^2})_f=
-{1\over 2}({X_g\over X_f})_{t,f}\label{AM1}
\ee
In the process of the evalution above we have used equations
connected with
derivatives of the functions $f$ and $X$(\ref{DDX}) and
(\ref{!}).
In (\ref{AM1}) the derivatives with respect to the argument
$t$ is only a
partial one. To have the possibility of taking it out of the
integration
sign of it is necessary to increase it to a total
derivative. This
is achieved by the following obvious manipulation:
$$
({X_g\over X_f})_{t,f}=\frac{d}{d t}({X_g\over
X_f})_f-({X_g\over X_f})_{f,f}
f_t
$$
Let us consider the possibility that the second term of the
last equality
in its turn is the total derivatives with respect to the
argument $t$. It is
obvious that for this it is sufficient assume additionally:
\be
{X_g\over X_f}=U(f,g)+a(t,g)f+b(t,g)\label{!!}
\ee
( but not the $t$ variable).

Under the assumptions above we can once integrate
(\ref{VBNF}) with the result:
\be
X_t+{1\over 2} \int^{g_2}_{g_1} dg g a(t,g)=P(g,f),\quad
X=\Phi(t)+P(g,f) t+
Q(g,f)\label{0}
\ee
from equations (\ref{!}) and (\ref{!!}) we obtain
additionally:
\be
X_f=(-(a_tf+b_t))^{-{1\over 2}},\quad
X_g=(-(a_tf+b_t))^{-{1\over 2}}(U+af+b)
\label{!!!})
\ee
The most simple way to resolve the last system of equations
is to consider as
a first step the condition of equality of the second mixed
partial derivatives and
compare one of the (selfconsistent) equations (\ref{!!!})
with (\ref{0}).

The condition of selfconsistency of the second mixed
derivatives reads as:
\be
a_{tg}f+b_{tg}=a_t(U+af+b)-2(U_f+a)(a_tf+b_t)\label{I}
\ee
In next three lines is presented the result of consequent
differentiation of
(\ref{I}) with the respect to the argument $f$:
$$
a_{tg}=-a_t(U_f+a)-2U_{ff}(a_tf+b_t)
$$
$$
0=3a_t{U_{ff}\over U_{fff}}+2(a_tf+b_t)
$$
$$
0=a_t(3({U_{ff}\over U_{fff}})_f+2)
$$
Going in the reverse direction we obtain in consequence (
all big letters are
arbitrary functions of argument $g$ only):
$$
U_{ff}=B(A-2f)^{-{3\over 2}},\quad U_f=B(A-2f)^{-{1\over
2}}+D,\quad
U=-B(A-2f)^{{1\over 2}}+Df+E
$$
$$
a_{tg}=-(D+a)a_t,\quad 2b+Aa=F
$$
The last equation for the function  $a$ after integration
once with  respect
to the argument $t$ may go over to the form ($a+D=z$) of a
Riccati
equation (now with respect to the argument $g$:
$$
z_g+{z^2\over 2}=\nu_g+{\nu^2\over 2}
$$
In the last equation we have represented the arbitrary
function arising in a special
form allowing us to integrate the Riccati equation with the
final result:
$$
a+D= -{\nu_{gg}\over \nu_g}+{2\nu_g\over \Phi(t)+\nu}
$$
After these preliminary calculations we are able to compare
the first equation
(\ref{!!!}) with (\ref{0}). For $(a_tf+b_t)$ from the
results above we have:
$$
(a_tf+b_t)=(A-2f){\nu_g\Phi_t\over (\Phi+\nu)^2}
$$
or
$$
(X_f)^2=(P_ft+Q_f)^2={(\Phi(t)+\nu)^2\over (2f-A)
\nu_g\Phi_t}
$$
All other computatrions are obvious and lead finally
precisely  to the solution
of the second subsection in section 3.

\end{document}